\documentclass[pra,tightenlines,twocolumn,showpacs,nofootinbib]{revtex4}
\usepackage{dcolumn,graphicx}
\begin{document}
\preprint{UNR Mar 2001-\today }
\title{ High-precision determination of transition amplitudes
of principal transitions in Cs from van der Waals coefficient $C_6$}

\author{Andrei Derevianko\email{andrei@unr.edu}}
\affiliation { Department of Physics, University of Nevada, Reno,
Nevada 89557}

\date{\today}

\begin{abstract}
A method for determination of atomic dipole matrix elements
of principal transitions from the value of dispersion coefficient $C_6$
of molecular potentials correlating to two ground-state atoms is proposed.
The method is illustrated
on atomic Cs using $C_6$ deduced from high-resolution Feshbach spectroscopy.
The following reduced matrix elements are determined
$\langle 6S_{1/2} || D || 6P_{1/2} \rangle =4.5028(60)\, |e| a_0 $ and
$\langle 6S_{1/2} || D || 6P_{3/2} \rangle =6.3373(84)\, |e| a_0 $
($a_0= 0.529177 \times 10^{-8}$ cm.) These matrix elements
are consistent with the results of the most accurate direct lifetime measurements
and have a similar uncertainty. It is argued that the uncertainty can be considerably
reduced as the coefficient $C_6$ is  constrained further.
\end{abstract}

%32.10.Dk, 34.50.Dy, 31.15.Ar

\maketitle
The leading long-range interaction of two atoms in their respective spherically-symmetric
ground states is described in terms of van der Waals coefficient $C_6$~\cite{DalDav66}.
Studies of magnetic-field induced Feshbach resonances and photoassociation in  ultracold
atomic samples allow to constrain this coefficient. Several highly-accurate determinations
of $C_6$ for alkali-metal dimers were recently
reported~\cite{AbeVer99,LeoWilJul00,DraTolTja00,RobBurCla01,VogFreTsa00}.
Here we propose a method for determination of electric-dipole matrix elements
(or lifetimes) of principal transitions from these coefficients.

The method is illustrated for atomic Cs. In recent years the most accurate
lifetimes for alkali-metal atoms were derived using photoassociation spectroscopy
of ultracold  atomic samples reviewed in Refs.~\cite{WeiBagZil99,StwWan99}.
Unfortunately, this approach was not as successful for Cs
because of peculiarities of molecular potentials of the Cs dimer~\cite{FioComDra99}.
We consider an alternative method.
Based on the van der Waals coefficient $C_6$ deduced from ultracold collision data,
the calculations reported here lead to the matrix elements for Cs with
an uncertainty similar to the best
direct lifetime measurements. We note that the determined high-precision
dipole matrix element for $6S_{1/2}-6P_{1/2}$ transition in Cs~\cite{melnote} is crucial
for an interpretation
of the most accurate measurement~\cite{WooBenCho97} of atomic parity nonconservation (PNC).
Atomic PNC~\cite{Khr91,BouBou97} provides powerful constraints on
possible extensions (e.g. extra Z-bosons) to the standard model of elementary particles.

For Cs, \citet{LeoWilJul00} deduced $C_6=6890(35)$\footnote{Unless specified otherwise,
atomic units $\hbar=|e|=m_e=1$ are used throughout the paper.}
using results of
high-resolution Feshbach spectroscopy~\cite{ChiVulKer00}.
Recently more Feshbach resonances due to higher
angular momenta were identified by Chu and co-workers and this value has been constrained
further to $C_6 =  6859(25)$~\cite{LeoTieWilJul01}.

\begin{figure}
\centerline{\includegraphics*[scale=0.55]{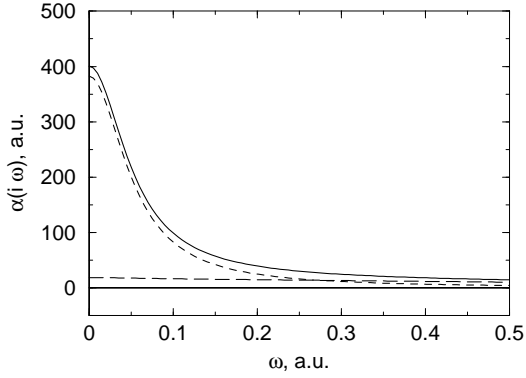}}
\caption{ Comparison of contributions to dynamic dipole polarizability
from the principal transitions $\alpha_{p}\left(  i\omega\right)$ (dashed line)
and residual intermediate states $\alpha_{r}\left(  i\omega\right)$ ( long-dashed line).
Total polarizability $\alpha\left(  i\omega\right)$ is represented by a solid line.
\label{Fig_alphas} }
\end{figure}

The dispersion coefficient $C_6$ characterizes a second-order dipole response
of atoms to molecular fields at large internuclear separations. It can be expressed as
a quadrature of dynamic polarizability~\cite{DalDav66}
\begin{equation}
C_{6}=\frac{3}{\pi}\int_{0}^{\infty}d\omega\,\,\left[  \alpha\left(
i\omega\right)  \right]  ^{2} \, ,
\end{equation}
where
\begin{equation}
\alpha\left(  i\omega\right)  =\frac{2}{3}\sum_{|i\rangle}\frac{\Delta E_{i}
}{\left( \Delta E_{i}\right)  ^{2}+\omega^{2}}\left|  \langle v|D|i\rangle
\right|  ^{2} \, .
\end{equation}
Here $|v\rangle$ is an atomic ground state ($6S_{1/2}$ for Cs), $\Delta E_{i}$ are energies of
intermediate states $| i \rangle$ taken with respect to the ground state, and
$D$ are electric-dipole matrix elements. For Cs almost 85\% of $C_6$ is
accumulated from intermediate states $6P_{1/2}$ and  $6P_{3/2}$.
We exploit this strong dependence to deduce matrix elements of principal
transitions by calculating
residual contributions using {\em ab initio} methods.
We separate the contribution of $6P_J$ states to dynamic polarizability
and write
\[
\alpha\left(  i\omega\right)  = \alpha_{p}\left(  i\omega\right)  +\alpha
_{r}\left(  i\omega\right) \, ,
\]
where $\alpha_{r}$ combines contributions of other intermediate states.
These polarizabilities are shown in Fig.~\ref{Fig_alphas}.
Introducing reduced matrix elements $D_{J}=\langle 6S_{1/2}||D||6P_{J}\rangle$
and a ratio  $R=(D_{3/2}/D_{1/2})^{2}$,
\begin{equation}
\alpha_{p}\left(  i\omega\right)  =\frac{1}{3}D_{1/2}^{2}\left(  \frac{\Delta
E_{1/2}}{\Delta E_{1/2}^{2}+\omega^{2}}+\frac{\Delta E_{3/2}}{\Delta
E_{3/2}^{2}+\omega^{2}}R\right) \, . \label{Eq_alphap}
\end{equation}
Energies $\Delta E_{J}$ of $6P_J$ states are known experimentally with a high accuracy and
the ratio of matrix elements $R=1.9809(9)$ has been measured with a 0.05\%
precision by \citet{RafTan98}. This accuracy is much better than the errors
in individual matrix elements due to a cancellation of systematic uncertainties.

\begin{figure}
\centerline{\includegraphics*[scale=0.65]{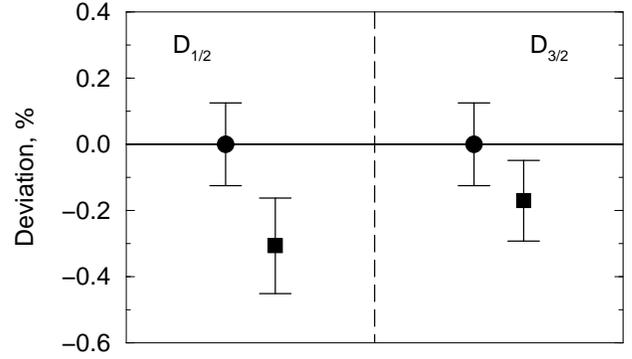}}
\caption{ Comparison of deduced (filled circles) values for matrix elements
$\langle 6P_{1/2} ||D||6S_{1/2}\rangle$ and $\langle 6P_{3/2} ||D||6S_{1/2}\rangle$
with the results of direct lifetime measurements~\protect\cite{RafTanLiv99} (filled squares). The values
are normalized to the present results and the deviation in \% is shown.
\label{Fig_comp} }
\end{figure}

Since
\begin{equation}
C_{6}=\frac{3}{\pi}\int_{0}^{\infty}d\omega\,\,\left[
\alpha_p\left(i\omega\right)^2 +
2 \alpha_p\left(i\omega\right)\alpha_r\left(i\omega\right)
+ \alpha_r\left(i\omega\right)^2
  \right]  \, ,
\end{equation}
the van der Waals coefficient can be parameterized in terms of the matrix element $D_{1/2}$ as
\begin{equation}
C_{6}=D_{1/2}^{4}\xi_{p}+D_{1/2}^{2}\xi_{x}+\xi_{r} \, ,
\label{Eqn_quadratic}
\end{equation}
where
\begin{eqnarray}
\xi_{p} &  = &\frac{1}{12}\left(  \frac{1}{\Delta E_{1/2}}+\frac{4R}{\Delta
E_{1/2}+\Delta E_{3/2}}+\frac{R^{2}}{\Delta E_{3/2}}\right) \, ,  \\
\xi_{x} &  =& \frac{2}{\pi}\int_{0}^{\infty}d\omega\,
\alpha_{r}\left(i\omega\right)  \times \nonumber \\
&&\left(  \frac{\Delta E_{1/2}}{\Delta E_{1/2}^{2}+\omega^{2}%
}+\frac{\Delta E_{3/2}}{\Delta E_{3/2}^{2}+\omega^{2}}R\right)  \label{Eq_xix} \, , \\
\xi_{r} &  =& \frac{3}{\pi}\int_{0}^{\infty}d\omega\,\,\left[  \alpha_{r}\left(
i\omega\right)  \right]  ^{2} \, .
\end{eqnarray}
Solving the quadratic equation~(\ref{Eqn_quadratic}) we obtain
\begin{equation}
D_{1/2}^{2}=\sqrt{\left(  \frac{C_{6}-\xi_{r}}{\xi_{p}}\right)  +\left(
\frac{\xi_{x}}{2\xi_{p}}\right)  ^{2}}-\frac{\xi_{x}}{2\xi_{p}}
\label{Eq_D}
\end{equation}
and the problem is reduced to an accurate determination of parameters $\xi$.
The calculation of these quantities and uncertainty estimates are discussed below.
We find $\xi_p = 14.0787$, $\xi_x = 46.05(92)$, and $\xi_r= 138.0(2.8)$
(the errors in the ratio $R$ will be treated separately) and obtain
\begin{eqnarray}
\langle 6P_{1/2} ||D||6S_{1/2}\rangle &=& 4.5028(60) \, ,\label{Eq_res}  \\
\langle 6P_{3/2} ||D||6S_{1/2}\rangle &=& 6.3373(84) \, .
\end{eqnarray}
These matrix elements are consistent with the
results of direct lifetime measurements~\cite{RafTanLiv99}
$\langle 6P_{1/2} ||D||6S_{1/2}\rangle = 4.4890(65)$ and
$\langle 6P_{3/2} ||D||6S_{1/2}\rangle = 6.3265(77)$ and have
a comparable accuracy (see Fig.~\ref{Fig_comp}).

{\em Details of calculation. ---}
The parameter $\xi_{p}$ can be evaluated with a high precision using experimental
energies and the ratio of matrix elements $R$ from Ref.\cite{RafTan98}.
We obtain   $\xi_{p}=14.0787$.
The uncertainty induced in this quantity
by experimental error in the ratio $R$ will be addressed separately.

To determine parameters $\xi_{x}$ and $\xi_{r}$ we have to compute the
residual dynamic polarizability $\alpha_r(i \omega)$. Here  we
follow formalism laid out in Ref.~\cite{DerJohSaf99} and augment it
with a refined error analysis.
The intermediate states can be separated into two classes - valence
states, both bound and continuum ($\alpha'_v$) , and core-excited states ($\alpha_c$)
\[
 \alpha_r\left(  i\omega\right) =
 \alpha'_v\left(  i\omega\right) +
 \alpha_c\left(  i\omega\right) +
 \alpha_{cv}\left(  i\omega\right) \, .
\]
Here term $\alpha_{cv}$ contains a small core-valence coupling correction
addressed below.

%First we review the overall calculation of the residual
%polarizability $\alpha_r\left(  i\omega\right)$ and than we
In summation over valence states $\alpha'_v$ we use a combination of relativistic
linearized coupled-cluster singles-doubles (CCSD) method and Dirac-Hartree-Fock (DHF) approximation.
In particular, the contribution from $7P_J$ and $8P_J$ states
is computed with the  matrix elements obtained in ~\cite{SafJohDer99}
and experimental energies~\cite{Moo58}.
The rest of the valence states is incorporated
using B-spline quasispectrum~\cite{JohBluSap88} generated in the
``frozen'' core DHF approximation.
%This procedure is numerically highly accurate and the error in $\alpha'_v$ is associated
%only with the omitted correlations.
In the summation over core-excited states we
employ relativistic random-phase approximation (RPA), described in Refs.~\cite{AmuChe75,Joh88} with
an obvious extension for frequency-dependence. In the RPA we allow for excitations of core
electrons
to all possible valence states, including the occupied $6S$ state. To account for
a subsequent violation of the Pauli exclusion principle, a counter core-valence coupling
term $\alpha_{cv}$ was introduced; we calculate it in the DHF approximation.

Now we proceed to the calculation of parameter $\xi_x$.
We notice that in Eq.~(\ref{Eq_xix})
a smooth and broad curve $\alpha_{r}\left(  i\omega\right)  $ is integrated with
a narrow Lorentzian-shaped function $\alpha_{p}\left(  i\omega\right)$
(see Fig.~\ref{Fig_alphas}).
To illustrate the main source of uncertainty we
approximate
\begin{equation}
\xi_x   \approx \alpha
_{r}\left(  0\right)  \frac{6}{\pi}\int_{0}^{\infty}\,
\frac{ \alpha_{p}\left(
i\omega\right)}{D_{1/2}^{2}}  d\omega
  =\alpha_{r}\left(  0\right)  (1+R) \, ,
\label{Eq_xixApprox}
\end{equation}
i. e. the uncertainty in $\xi_x$ is governed by static
residual polarizability $\alpha_{r}\left(  0\right)$.
We present a breakdown of various contributions to this polarizability
in Table~\ref{Tab_alhaR0}.

\begin{table}
\caption{ Breakdown of various {\em ab initio} contributions to
static dipole polarizability from the intermediate states beyond $6P_J$.
\label{Tab_alhaR0}}
\begin{ruledtabular}
\begin{tabular}{llll}
 &
 States &
\multicolumn{1}{c}{Value } &
\multicolumn{1}{c}{Method}  \\
\hline

$\alpha_v(0)$   &$7P_J$             &  1.50  &CCSD\footnotemark[1]\\
                &$8P_J$             &  0.18  &CCSD\footnotemark[1] \\
                &$9\cdots\infty P_J$&  0.33  &DHF \\
                & Total             &  2.01  & \\
$\alpha_{cv}(0)$&                   & -0.47  &DHF \\
$\alpha_c(0)$   &                   &  15.81 &RRPA \\
\hline
$\alpha_r(0)$   &                  &   17.35 &      \\
\end{tabular}
\end{ruledtabular}
\footnotetext[1]{ {\em ab initio} matrix elements and experimental energies.}
\end{table}

The error bars of the derived matrix elements depend sensitively on
the uncertainty of $\xi_x$.
To estimate this uncertainty,
it is instructive to discuss characteristic accuracies of various
relativistic many-body methods. As a test case we consider
contribution of principal transitions to the static dipole polarizability
\begin{equation}
\alpha_p(0)  = \frac{1}{3} \left(
\frac{D_{1/2}^2}{\Delta E_{1/2}} + \frac{D_{3/2}^2}{\Delta E_{1/2}}
\right) \, . \label{Eq_alphap0}
\end{equation}
We calculate this quantity in (i) Dirac-Hartree-Fock (DHF) approximation,
(ii) linearized coupled-cluster singles-doubles (CCSD) method~\cite{SafJohDer99},
and (iii) using experimental energies~\cite{Moo58} and matrix elements~\cite{RafTanLiv99}.
The results are presented in Table~\ref{Tab_test}. It is clear that the
Hartree-Fock approximation has a 70\% error and the CCSD values
have an accuracy in the order of 1\%. Based on this example we assign a 1\%
uncertainty to CCSD contributions from $7P_J$ and $8P_J$ intermediate
states and a 70\% uncertainty to the corrections from the
rest of the valence state and 70\% to core-valence coupling term.
We further replace the RPA value
for the static polarizability of the core with accurate semiempirical
value 15.644(5)~\cite{ZhoNor89} and obtain $\alpha^{\rm s.e.}_r(0) =17.18(40)$.
Almost entire error comes from the uncertainty of the Hartree-Fock method.
It is worth noting that  the overall accuracy of the $\alpha_v(0)$ and
the derived matrix elements can be further improved using, for example,
linear-response coupled-cluster method~\cite{Liu89}.
Based on Eq.~(\ref{Eq_xixApprox}) we improve the accuracy
of the parameter $\xi_x$ by rescaling the {\em ab initio} value
of the integral, Eq.~(\ref{Eq_xix}), with the semiempirical $\alpha^{\rm s.e.}_r(0)$. The final value
for the quantity $\xi_x$ is 46.0(1.0) .

\begin{table}
\caption{ Characteristic accuracy of {\em ab initio} methods in calculations
of contribution of principal transitions to static polarizability of Cs ground
state.
Accuracy of DHF and CCSD methods is defined with respect to the experimental value (third row).
\label{Tab_test}}
\begin{ruledtabular}
\begin{tabular}{ldl}
Method &
\multicolumn{1}{c}{ Value } &
\multicolumn{1}{c}{Accuracy}  \\
\hline
DHF   & 644     &  70\% \\
CCSD\footnotemark[1]
      & 378.6  & 0.7\%  \\
Expt. & 381.2  & 0.2\%\footnotemark[2] \\
\end{tabular}
\end{ruledtabular}
\footnotetext[1]{ {\em ab initio} matrix elements and experimental energies.}
\footnotetext[2]{ Based on experimental error bars in matrix elements from
Ref.~\protect\cite{RafTanLiv99}}
\end{table}

\begin{figure}
\centerline{\includegraphics*[scale=1.00]{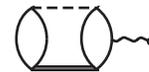}}
\caption{ Sample Brueckner-Goldstone diagram
illustrating coupling of a doubly excited core state (double horizontal line)
to the many-body ground state of a closed-shell core
through a one-body dipole matrix element (wavy line).
Such contribution is possible only in a combination  with correlation part of the
wavefunction of the ground state
(first order correction is shown, with Coulomb interaction represented by dashed line.)
\label{Fig_diag} }
\end{figure}

The overwhelming contribution to parameter $\xi_r$ comes from the dynamic core polarizability
$\alpha_c\left(  i\omega\right)$.
In summation over core-excited states we employ relativistic random-phase
approximation (RPA), described in Refs.~\cite{AmuChe75,Joh88} with
an obvious extension for frequency-dependence. Because of the equivalence
of length- and velocity-forms of dipole matrix elements
in RPA, the calculated dynamic polarizability of the core
satisfies an important Thomas--Reiche--Kuhn (TRK) sum rule $\lim_{\omega \rightarrow \infty}
\alpha_c(i \omega) = N_c/\omega^2$, $N_c$ being the number of core electrons
($N_c = 54$ for Cs.) This property is illustrated in Fig.~1 of Ref.~\cite{DerBabDal01}.
The static ($\omega=0$) core polarizability obtained with RPA is $\alpha_c(0) =15.81$.
This result is in a 1\% agrement with a value of 15.644(5)  deduced
from semiempirical fitting of Rydberg spectrum of Cs~\cite{ZhoNor89}.
The RPA accounts only for a correlated contribution of
particle-hole excitations. However, the correction from multiple core excitations
is expected to be strongly suppressed because they can couple only to the
first-order many-body correction to the core wavefunction (see Fig.~\ref{Fig_diag}). In addition,
the effect of multiple core excitations is reduced  by larger energy denominators in
the expression for polarizability.

Overall the RPA approximation results in a dynamic
core polarizability differing from the high-precision value by 1\% at $\omega=0$
and at large frequencies recovering the exact TRK limit~\cite{TRKComment}.
The parameter $\xi_r$ calculated including only core polarizability,
$\xi_r=137.2$, is slightly increased by inclusion of valence states and the
core-valence coupling term $\alpha_{cv}$ to 138.0. Based on the above discussion we assign a 2\% uncertainty to $\xi_r$ and
arrive at $\xi_r = 138.0(2.8)$. It should be noted that  more conservative
error bar of 5\% for $\xi_r$ does not significantly affect the resulting
precision of derived matrix elements.

Employing $C_6 =  6859(25)$~\cite{LeoTieWilJul01} and
the calculated parameters $\xi$ we determine the matrix element $D_{1/2}$, and,
using the ratio $R$~\cite{RafTan98},
the matrix element $D_{3/2}$. The obtained
values are given in Eq.~(\ref{Eq_res}). The uncertainty in matrix element
calculated from Eq.~(\ref{Eq_D}) can be parameterized as
\begin{eqnarray*}
\lefteqn{ \left(  \frac{\delta D_{1/2}}{D_{1/2}}\right)  ^{2} =
A_{C_{6}}\, \left(
\frac{\delta C_{6}}{C_{6}}\right)  ^{2} +} \\
& + &A_{x} \, \left(  \frac{\delta\xi_{x}
}{\xi_{x}}\right)  ^{2}+A_{r}\, \left(  \frac{\delta\xi_{r}}{\xi_{r}
}\right)  ^{2}+A_{R}\, \left(  \frac{\delta R}{R}\right)  ^{2} \, ,
\end{eqnarray*}
where $A_{C_{6}}=7.5\times10^{-2},\,A_{r}=3.0\times10^{-5},A_{x}%
=1.4\times10^{-3}$, and $A_{R}=9.4\times10^{-2}$. Combining estimated uncertainties
we obtain an error bound of 0.13\% for matrix elements. This accuracy
is similar to that of the best direct lifetime measurements by \citet{RafTanLiv99}.
The reader is referred to Ref.~\cite{RafTanLiv99} for an extensive comparison
with other measurements and theoretical predictions.
Our result is most sensitive to the errors in the van der Waals coefficient
and parameter $\xi_x$. Provided that $C_6$ is known exactly, the uncertainty
in the matrix elements can be reduced to 0.07\%  with the current technique
for estimation of $\xi_x$. It is worth noting that the precision of calculation for $\xi_x$
can be substantially improved, for example, using linear-response
coupled-cluster method~\cite{Liu89}.

{\em Conclusion.---}
We exploited a strong dependence of the van der Waals
coefficient $C_6$ on matrix elements of principal
transitions.  We deduced these matrix elements by calculating
small residual contributions using {\em ab initio} methods.
The proposed method was applied to Cs atom and the derived
matrix elements are consistent with the best direct lifetime
measurements~\cite{RafTanLiv99} and have a similar uncertainty.

In anticipation of high-precision measurements of static dipole
polarizability $\alpha(0)$ of the ground state of Cs with atom
interferometry~\cite{EksSchCha95},
we note that 96\% of the polarizability is due to the contribution of
the principal transitions $\alpha_p(0)$~\cite{DerJohSaf99}. Subtracting
the residual contribution $\alpha^{\rm s.e.}_r(0) =17.18(40)$ from
the measured $\alpha(0)$ one can also determine matrix elements
of principal transitions to an accuracy of 0.05\% from Eq.~(\ref{Eq_alphap}).

We employed the 0.36\%-accurate value of the van der Waals coefficient $C_6$
deduced by \citet{LeoTieWilJul01} from high-resolution Feshbach spectroscopy
of ultracold Cs atoms~\cite{ChiVulKer00}.
According to Ref.~\cite{ChiVulKer00} the precision of this dispersion coefficient
can be potentially improved to $0.03\%$.
If such a precision is achieved, the method proposed here, augmented
with more accurate {\em ab initio} calculations and better measurements of
the ratio $R$, can lead to determination of
matrix elements of principal transitions with an unprecedented 0.01\% accuracy.

I would like to thank C.J. Williams and P.S. Julienne for communicating
the updated value of $C_6$ coefficient~\cite{LeoTieWilJul01} for Cs dimer
prior to publication.
Thanks are also due to D.E. Pritchard and A.D. Cronin for discussions  of
prospects of high-precision measurement of Cs static polarizability.
This work was partially supported by the National Science Foundation.

\bibliography{mypub,vdW,general,lifetimes,exact,pnc,notes,rpa}

\end{document}